\documentclass{article}
\usepackage{spconf,amsmath,graphicx}
\usepackage{hyperref}

\usepackage{booktabs}
\usepackage{xcolor,colortbl}
\usepackage{tikz}
 
\newcommand{\myparagraph}[1]{\vspace{.4em} \noindent \textbf{#1}\ }
\usepackage[linesnumbered,ruled,vlined]{algorithm2e}
\newcommand\crule[3][black]{\textcolor{#1}{\rule{#2}{#3}}}

\SetCommentSty{mycommfont}

\SetKwInput{KwInput}{Input}                
\SetKwInput{KwOutput}{Output}              
\SetKwInput{KwInit}{Given}              

\title{
On the Interplay between Sparsity, Naturalness, \\Intelligibility, and Prosody in Speech Synthesis}

\name{\begin{tabular}{c}
Cheng-I Jeff Lai\textsuperscript{1,2}, Erica Cooper\textsuperscript{* 3}, Yang Zhang\textsuperscript{* 2}, Shiyu Chang\textsuperscript{2}, Kaizhi Qian\textsuperscript{2}, Yi-Lun Liao\textsuperscript{1}\\ Yung-Sung Chuang\textsuperscript{1}, Alexander H. Liu\textsuperscript{1}, Junichi Yamagishi\textsuperscript{3}, David Cox\textsuperscript{2}, James Glass\textsuperscript{1}
\thanks{\hspace{-1mm}* EC and YZ contribute equally.}
\end{tabular}}

\address{\textsuperscript{1}MIT CSAIL, USA\quad \textsuperscript{2}MIT-IBM Watson AI Lab, USA\quad \textsuperscript{3}National Institute of Informatics, Japan\\
\small \texttt{\href{clai24@mit.edu}{clai24@mit.edu}}}

\usepackage{amssymb}
\usepackage{pifont}
\usepackage{amsmath,bm,multicol,multirow}
\usepackage{tablefootnote}
\usepackage{longtable}

\begin{document}
\ninept
\maketitle

\vspace{0mm}
\begin{abstract}
\vspace{-0mm}
Are end-to-end text-to-speech (TTS) models over-parametrized?
To what extent can these models be pruned, and what happens to their synthesis capabilities?
This work serves as a starting point to explore pruning both spectrogram prediction networks and vocoders. 
We thoroughly investigate the tradeoffs between sparsity and its subsequent effects on synthetic speech.
Additionally, we explore several aspects of TTS pruning: amount of finetuning data versus sparsity, TTS-Augmentation to utilize unspoken text, and combining knowledge distillation and pruning. 
Our findings suggest that not only are end-to-end TTS models highly prunable, but also, perhaps surprisingly, pruned TTS models can produce synthetic speech with equal or higher naturalness and intelligibility, with similar prosody. 
All of our experiments are conducted on publicly available models, and findings in this work are backed by large-scale subjective tests and objective measures. 
Code and 200 pruned models are made available to facilitate future research on efficiency in TTS\footnote{Project webpage: \url{https://people.csail.mit.edu/clai24/prune-tts/}}.
\end{abstract}


\vspace{-3mm}
\section{Introduction}
\label{sec:intro}
\vspace{-2mm}
End-to-end text-to-speech (TTS)\footnote{We refer to end-to-end TTS systems as those composed of an acoustic model (also known as text-to-spectrogram prediction network) and a separate vocoder, as there are relatively few direct text-to-waveform models; see~\cite{tan2021survey}.} research has focused heavily on modeling techniques and architectures, aiming to produce more natural, adaptive, and expressive speech in robust, low-resource, controllable, or online conditions~\cite{tan2021survey}.
We argue that an overlooked orthogonal research direction in end-to-end TTS is \textit{architectural efficiency}, and in particular, there has not been any established study on pruning end-to-end TTS in a principled manner. 
As the body of TTS research moves toward the mature end of the spectrum, we expect a myriad of effort delving into developing efficient TTS, with direct implications such as on-device TTS or a better rudimentary understanding of training TTS models from scratch~\cite{frankle2018lottery}. 

To this end, we provide analyses on the effects of pruning end-to-end TTS, utilizing basic unstructured magnitude-based weight pruning\footnote{Given that there has not been a dedicated TTS pruning study in the past, we resort to the most basic form of pruning. For more advanced pruning techniques, please refer to~\cite{gale2019state,blalock2020state}.}.
The overarching message we aim to deliver is two-fold: 
\begin{itemize}
    \item End-to-end TTS models are over-parameterized; their weights can be pruned with unstructured magnitude-based methods.
    \item Pruned models can produce synthetic speech at equal or even better naturalness and intelligibility with similar prosody.
\end{itemize}

\noindent To introduce our work, we first review two areas of related work: 
\vspace{-1mm}
\myparagraph{Efficiency in TTS}
One line of work is on small-footpoint, fast, and parallelizable versions of WaveNet~\cite{oord2016wavenet} and WaveGlow~\cite{prenger2019waveglow} vocoders; prominent examples are WaveRNN\footnote{Structured pruning was in fact employed in WaveRNN, but merely for reducing memory overhead for the vocoder. What sets this work apart is our pursuit of the scientific aspects of pruning end-to-end TTS holistically.}~\cite{kalchbrenner2018efficient}, WaveFlow~\cite{ping2020waveflow}, Clarinet~\cite{ping2018clarinet}, HiFi-GAN~\cite{kong2020hifi}, Parallel WaveNet~\cite{oord2018parallel}, SqueezeWave~\cite{zhai2020squeezewave}, DiffWave~\cite{kong2020diffwave}, WaveGrad 1~\cite{chen2020wavegrad}, Parallel WaveGAN~\cite{yamamoto2020parallel} etc. 
Another is acoustic models based on non-autoregressive generation (ParaNet~\cite{peng2020non}, Flow-TTS~\cite{miao2020flow}, MelGAN~\cite{kumar2019melgan}, EfficientTTS~\cite{miao2021efficienttts}, FastSpeech~\cite{ren2019fastspeech,ren2020fastspeech}), neural architecture search (LightSpeech~\cite{luo2021lightspeech}), diffusion (WaveGrad 2~\cite{chen2021wavegrad}), etc.
Noticeably, efficient music generation has gathered attention too, e.g. NEWT~\cite{hayes2021neural} and DDSP~\cite{engel2020ddsp}.

\begin{figure*}[ht]
\centering
\includegraphics[width=0.95\linewidth]{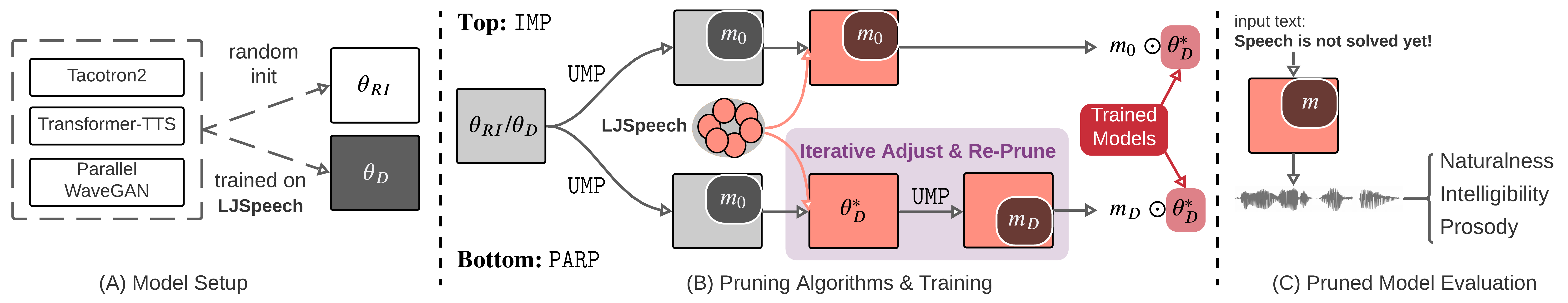}
\vspace{-4mm}
\caption{Illustration of our end-to-end TTS pruning setup.
\textbf{Left:} three TTS models are considered: Tacotron2, Transformer-TTS, and Parallel WaveGAN. By default, we set the initial weight $\theta_{0}$ to trained models $\theta_{D}$ on LJSpeech, but they can also be randomly initialized $\theta_{RI}$. 
\textbf{Middle:} top row is the ${\tt IMP}$ Baseline, and bottom row is ${\tt PARP}$.
Both are architecture-agnostic, and utilize ${\tt UMP}$ for retrieving initial pruning mask $m_{0}$. 
The only difference is $m_{0}$ is adjustable in ${\tt PARP}$ during training, while being fixed in ${\tt IMP}$.
Both algorithms produce pruned subnetworks $m\odot\theta_{D}^*$ that are finetuned on LJSpeech. 
\textbf{Right:} we evaluate pruned model synthetic speech's naturalness, intelligibility, and prosody via large-scale subjective and objective tests across sparsities.
}
\vspace{-5mm}
\label{fig:illustration}
\end{figure*}

\vspace{-1mm}
\myparagraph{ASR Pruning}
Earlier work on ASR pruning reduces the FST search space, such as~\cite{xu2018pruned}. 
More recently, the focus has shifted to pruning end-to-end ASR models~\cite{yu2012exploiting,shangguan2019optimizing,wu2021dynamic,lai2021parp}. 
Generally speaking, pruning techniques proposed for vision models~\cite{gale2019state,blalock2020state} work decently well in prior ASR pruning work, which leads us to ask, how effective are simple pruning techniques for TTS? 

\vspace{1mm}
This work thus builds upon a recent ASR pruning technique termed ${\tt PARP}$~\cite{lai2021parp},  with the intention of not only reducing architectural complexity for end-to-end TTS, but also demonstrating the surprising efficacy and simplicity of pruning in contrast to prior TTS efficiency work.
We first review ${\tt PARP}$ in Section~\ref{sec:method}.
In Section~\ref{sec:exp_setup}, we describe our experimental and listening test setups, and in Section~\ref{sec:results} we present results with several visualizations. 
Our contributions are: 

\vspace{-1mm}
\begin{itemize}
  \item We present the first comprehensive study on pruning end-to-end acoustic models (Transformer-TTS~\cite{li2019neural}, Tacotron2~\cite{shen2018natural}) and vocoders (Parallel WaveGAN~\cite{yamamoto2020parallel}) with an unstructured magnitude based pruning method ${\tt PARP}$~\cite{lai2021parp}.
  
  \item We extend ${\tt PARP}$ with knowledge distillation (${\tt KD}$) and TTS-Augmentation~\cite{hwang2021tts} for TTS pruning, demonstrating ${\tt PARP}$'s applicability and effectiveness regardless of network architectures or input/output pairs. 
  
  \item We show that end-to-end TTS models are over-parameterized. 
  Pruned models produce speech with similar levels of naturalness, intelligibility, and prosody to that of unpruned models. 
  
  \item For instance, with large-scale subjective tests and objective measures, Transformer-TTS at 30\% sparsity has statistically better naturalness than its original version; 
  for another, small footprint CNN-based vocoder has little to no synthesis degradation at up to 88$\%$ sparsity. 
  
\end{itemize}

\vspace{-2mm}
\section{Method}
\label{sec:method}
\vspace{-2.5mm}
\subsection{Problem Formulation}
\label{subsec:problem_formulation}
\vspace{-2mm}
Consider a sequence-to-sequence learning problem, where $\bm{X}$ and $\bm{Y}$ represent the input and output sequences respectively. 
For ASR, $\bm{X}$ is waveforms and $\bm{Y}$ is character/phone sequences; for a TTS acoustic model, $\bm{X}$ is character/phone sequences and $\bm{Y}$ is spectrogram sequences; for a vocoder, $\bm{X}$ is spectrogram sequences and $\bm{Y}$ is waveforms. 
A mapping function $f(\bm{X}; \theta)$ parametrized by a neural network is learned, where $\theta \in \mathcal{R}^d$ represents the network parameters and $d$ represents the number of parameters.
Sequence-level log-likelihood $\mathop{\mathbb{E}}\big[\ln P(\bm{Y}\mid\bm{X};\theta)]$ on target dataset $\mathcal{D}$ is maximized.

Our goal is to find a subnetwork $m \odot \theta$, where $\odot$ is the element-wise product and a binary pruning mask $m \in \{0, 1\}^d$ is applied on the model weights $\theta$. 
The ideal pruning method would learn $m$ at target sparsity such that $f(\bm{X}; m \odot \theta)$ achieves similar loss as $f(\bm{X};\theta)$ after training on $\mathcal{D}$.

\vspace{-2.5mm}
\subsection{Pruning Sequence-to-Sequence Models with ${\tt PARP}$}
\label{subsec:prune_seq2seq}
\vspace{-2.5mm}

\myparagraph{Unstructured Magnitude Pruning (${\tt UMP}$)}~\cite{frankle2018lottery,gale2019state}
sorts the model's weights according to their magnitudes across layers regardless of the network structure, and removes the smallest ones to meet a predefined sparsity level.
Weights that are pruned out (specified by $m$) are zeroed out and do not receive gradient updates during training. 

\myparagraph{Iterative Magnitude Pruning (${\tt IMP}$)}~\cite{frankle2018lottery,gale2019state}
is based on ${\tt UMP}$ and assumes an initial model weight $\theta_0$ and a target dataset $\mathcal{D}$ are given. ${\tt IMP}$ can be described as: 
\vspace{-1.5mm}
\begin{enumerate}
    \item Directly prune $\theta_0$ at target sparsity, and obtain an initial pruning mask $m_0$. Zero out weights in $\theta_0$ given by $m_0$. 
    \vspace{-1mm}
    \item Train $f(\bm{X}; m_0 \odot \theta_0)$ on $\mathcal{D}$ until convergence. Zeroed-out weights do not receive gradient updates via backpropogation. 
\end{enumerate}
\vspace{-1.5mm}
The above procedure can be iterated multiple times by updating $\theta_0$ with the finetuned model weight $\theta_D^*$ from Step 2. 

\vspace{-1mm}
\myparagraph{Prune-Adjust-Re-Prune (${\tt PARP}$)}~\cite{lai2021parp} 
is a simple modified version of ${\tt IMP}$ recently proposed for self-supervised speech recognition, showing that pruned wav2vec 2.0~\cite{baevski2020wav2vec} attains lower WERs than the full model under low-resource conditions.
Given its simplicity, here we show that ${\tt PARP}$ can be applied to any sequence-to-sequence learning scenario. 
Similarly, given an initial model weight $\theta_0$ and $\mathcal{D}$, ${\tt PARP}$ can be described as (See Fig~\ref{fig:illustration} for visualization):
\vspace{-1.5mm}
\begin{enumerate}
    \item Same as ${\tt IMP}$'s Step 1. 
    \vspace{-1mm}
    \item Train $f(\bm{X}; \theta_0)$ on $\mathcal{D}$. Zeroed-out weights in $\theta_0$ receive gradient updates via backprop. After $N$ model updates, obtain the trained model $f(\bm{X};\theta_D^*)$, and apply ${\tt UMP}$ on $\theta_D^*$ to obtain mask $m_{D}$. Return subnetwork $m_{D} \odot \theta_D^*$.
\end{enumerate}

\vspace{-1mm}
\myparagraph{Setting Initial Model Weight $\theta_0$}
In~\cite{lai2021parp}, ${\tt PARP}$'s $\theta_0$ can be the self-supervised pretrained initializations, or any trained model weight $\theta_P$ ($P$ needs not be the target task $D$). 
On the other hand, ${\tt IMP}$'s $\theta_0$ is target-task dependent i.e. $\theta_0$ is set to a trained weight on $\mathcal{D}$, denoted as $\theta_D$.
However, since the focus in this work is on the final pruning performance only, we set $\theta_0$ to $\theta_D$ by default for both ${\tt PARP}$ and ${\tt IMP}$.

\vspace{-0.5mm}
\myparagraph{Progressive Pruning with ${\tt PARP}$-${\tt P}$} 
Following~\cite{lai2021parp}, we also experiment with progressive pruning (${\tt PARP}$-${\tt P}$), where ${\tt PARP}$-${\tt P}$'s Step 1 prunes $\theta_0$ at a lower sparsity, and its Step 2 progressively prunes to the target sparsity every $N$ model updates. 
We show later that ${\tt PARP}$-${\tt P}$ is especially effective in higher sparsity regions. 

\vspace{-3.5mm}
\section{Experimental Setup}
\label{sec:exp_setup}
\vspace{-2mm}
\subsection{TTS Models and Data}
\label{subsec:model_and_data}
    \vspace{-3mm}
    \myparagraph{Model Configs}
    Our end-to-end TTS is based on an acoustic model (phone to melspec) and a vocoder (melspec to wav). 
    To ensure reproducibility, we used publicly available and widely adopted implementations\footnote{Checkpoints are also available at \href{https://github.com/espnet/espnet}{ESPnet} and \href{https://github.com/kan-bayashi/ParallelWaveGAN}{ParallelWaveGAN}.}: Transformer-TTS~\cite{li2019neural} and Tacotron2~\cite{shen2018natural} as the acoustic models, and Parallel WaveGAN~\cite{yamamoto2020parallel} as the vocoder. 
    Transformer-TTS and Tacotron2 have the same high-level structure (encoder, decoder, pre-net, post-net) and loss (l2 reconstructions before and after post-nets and stop token cross-entropy). 
    Transformer-TTS consists of a 6-layer encoder and a 6-layer decoder. Tacotron2's encoder consists of 3-layer convolutions and a BLSTM, and its decoder is a 2-layer LSTM with attention.
    Both use a standard G2P for converting text to phone sequences as the model input.
    Parallel WaveGAN consists of convolution-based generator $G$ and discriminator $D$.
    
    \vspace{-1mm}
    \myparagraph{Datasets}
    LJspeech~\cite{ljspeech17} is used for training acoustic models and vocoders. 
    It is a female single-speaker read speech corpus with 13k text-audio pairs, totaling 24h of recordings. 
    We also used the transcription of Librispeech's {\tt train-clean-100} partition~\cite{panayotov2015librispeech} as additional unspoken text\footnote{Both LJspeech and Librispeech are based on audiobooks.} used in TTS-Augmentation. 
    
\vspace{-4mm}
\subsection{${\tt PARP}$ Implementation}
\label{subsec:parp_implementation}
\vspace{-1.5mm}
    ${\tt UMP}$ is based on PyTorch's API\footnote{\href{https://pytorch.org/tutorials/intermediate/pruning_tutorial.html}{PyTorch Pruning API}}.
    For all models, $\theta_0$ is set to pretrained checkpoints on LJspeech, and $N$ is set to 1 epoch of model updates. 
    We jointly prune encoder, decoder, pre-nets, and post-nets for the acoustic model; for vocoder, since only $G$ is needed during test-time synthesis, only $G$ is pruned ($D$ is still trainable).
     
\begin{figure*}[!htbp]
\centering
\includegraphics[width=0.88\linewidth]{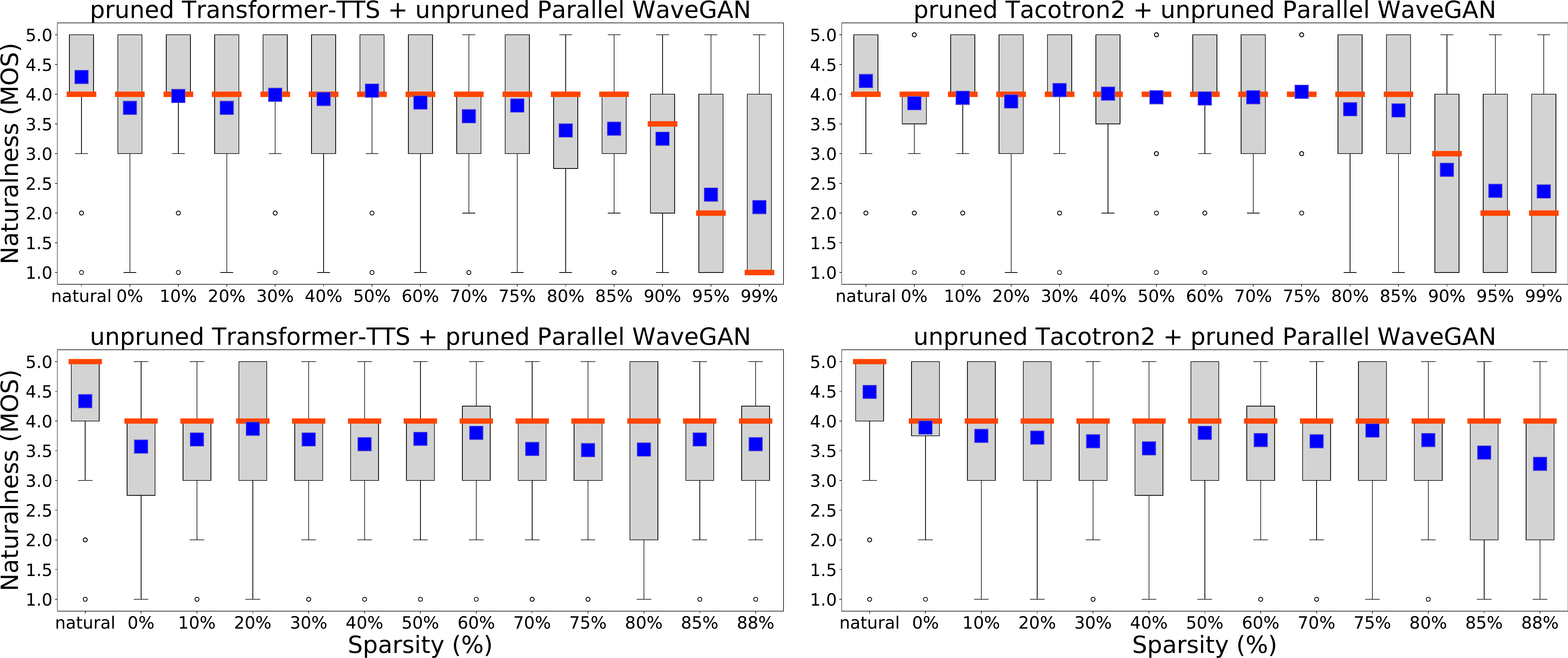}
\vspace{-3.5mm}
\caption{
Box plots for four independent MOS tests across configurations (pruned/unpruned acoustic models + pruned/unpruned vocoders).
At each sparstiy, \crule[blue]{0.2cm}{0.2cm} is the mean and \crule[red]{0.3cm}{0.07cm} is the median MOS score over 100 HITs. 
Ground truth recordings (natural) are included as the topline. 
}
\definecolor{mediumslateblue}{rgb}{0.48, 0.41, 0.93}
\label{fig:mos}
\vspace{-4.5mm}
\end{figure*}

\vspace{-3mm}
\subsection{Complementary Techniques for ${\tt PARP}$}
\label{subsec:additional_techniques}
\vspace{-3mm}
\myparagraph{TTS-Augmentation for unspoken transcriptions}
The first technique is based on TTS-Augmentation~\cite{hwang2021tts}.
It is a form of self-training, where we take $f(\theta_D)$ to label additional unspoken text $\bm{X}_u$.
The newly synthesized paired data, denoted $\mathcal{D}_u = (\bm{X}_u, f(\bm{X}_u;\theta_D))$, is used together with $\mathcal{D}$ in ${\tt PARP}$'s Step 2. 

\myparagraph{Combining Knowledge-Distillation (${\tt KD}$) and ${\tt PARP}$,} with a teacher model denoted as $f(\theta_D)$.
The training objective in ${\tt PARP}$'s Step 2 is set to reconstructing both ground truth melspec and melspec synthesized by an (unpruned) teacher acoustic model $f(\theta_D)$.

\vspace{-3mm}
\subsection{Subjective and Objective Evaluations}
\label{subsec:sub_and_obj_eval}
\vspace{-1.5mm}
We examine the following three aspects of the synthetic speech:   
\begin{itemize}

    \vspace{-0.5mm}
    \item \textbf{Naturalness} is quantified by the 5-point (1-point increment) scale Mean Opinion Score (MOS). 
    20 unique utterances (with 5 repetitions) are synthesized and compared across pruned models, for a total of 100 HITs (crowdsourced tasks) per MOS test. 
    In each HIT, the input texts to all models are the same to minimize variability.
    
    \vspace{-0.5mm}
    \item \textbf{Intelligibility} is measured with Google's ASR API\footnote{\url{https://pypi.org/project/SpeechRecognition/}}.
    
    \vspace{-0.5mm}
    \item \textbf{Prosody} via mean and standard deviation (std) fundamental frequency ($F_{0}$) estimations\footnote{$F_{0}$ estimation with probabilistic YIN (pYIN) implemented in \href{https://librosa.org/doc/main/generated/librosa.yin.html}{Librosa}.} and utterance duration, averaged over dev and eval utterances.
\end{itemize}

\noindent We also perform pairwise comparison (A/B) testings for naturalness and intelligibility (separately). 
Similar to our MOS test, we release 20 unique utterances (with 10 repetitions), for a total of 200 HITs per A/B test.
In each HIT, input text to models are also the same. 
MOS and A/B tests are conducted in Amazon Mechanical Turk (AMT).

\myparagraph{Statistical Testing}
To ensure our AMT results are statistically significant, we run Mann-Whitney U test for each MOS test, and pairwise z-test for each A/B test, both at significance level of $p\le 0.05$.


\vspace{-3mm}
\section{Results}
\label{sec:results}
\vspace{-2mm}


\vspace{-0mm}
\subsection{Does Sparsity improve Naturalness?}
\label{subsec:exp_naturalness}
\vspace{-1.5mm}
Fig~\ref{fig:mos} is the box plot of MOS scores of pruned end-to-end TTS models at 0\%$\sim$99\% sparsities with ${\tt PARP}$. 
In each set of experiments, only one of the acoustic model or vocoder is pruned, while the other is kept intact. 
For either pruned Transformer-TTS or Tacotron2 acoustic models, their MOS scores are statistically not different from the unpruned ones at up to 90\% sparsity.
For pruned Parallel WaveGAN, pairing it with an unpruned Transformer-TTS reaches up to 88\% sparsity without \textit{any} statistical MOS decrease, and up to 85\% if paired with an unpruned Tacotron2. 
Based on these results, we first conclude that end-to-end TTS models are over-parameterized across model architectures, and removing the majority of their weights does not significantly affect naturalness. 

Secondly, we observe that the 30\% pruned Tacotron2 has a  statistically higher MOS score than unpruned Tacotron2. 
Although this phenomenon is not seen in Transformer-TTS, WaveGAN, or at other sparsities, it is nonetheless surprising given ${\tt PARP}$'s simplicity.
We can hypothesize that under the right conditions, \textit{pruned models train better}, which results in higher naturalness over unpruned models. 

\vspace{-4.5mm}
\subsection{Does Sparsity improve Intelligibility?}
\label{subsec:exp_intelligibility}
\vspace{-5mm}

\begin{figure}[!bh]
\centering
\includegraphics[width=0.9\linewidth]{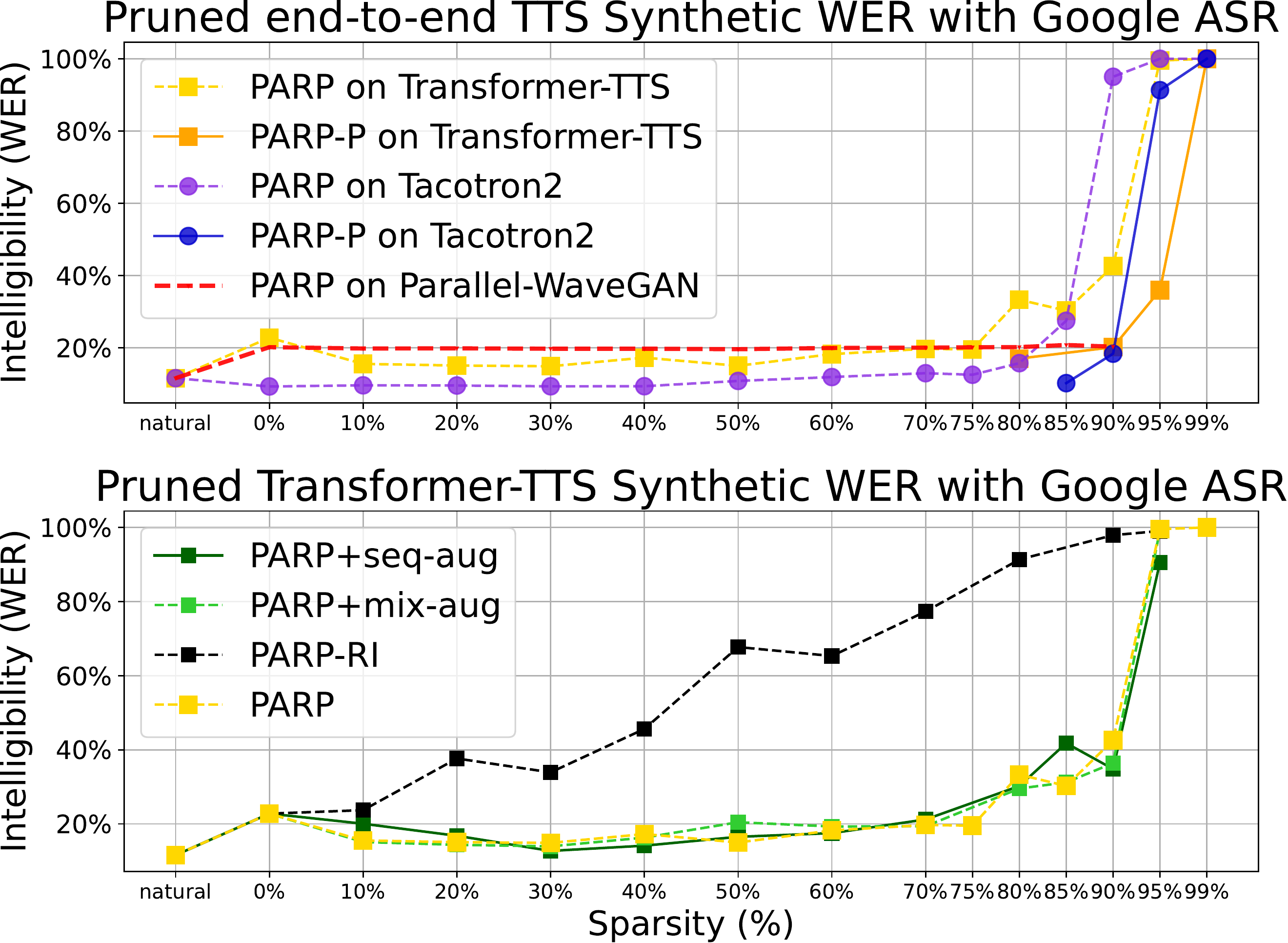}
\vspace{-4mm}
\caption{\textbf{Top} plots the synthetic speech WERs over sparsities for all model combinations.
\textbf{Bottom} compares the WERs for different pruning configurations.}
\label{fig:wer}
\vspace{-4mm}
\end{figure}

We measure intelligibility of synthetic speech via Google ASR, and Figure~\ref{fig:wer} plots synthetic speech's WERs across sparsities over model and pruning configurations. 
Focusing on the top plot, we have the following two high-level impressions: 
(1) WER decreases at initial sparsities and increases dramatically at around 85\% sparsity with ${\tt PARP}$ (yellow and purple dotted lines).  
(2) pruning the vocoder does not change the WERs at all (observe the straight red dotted line). 

Specifically, for Transformer-TTS, ${\tt PARP}$ at 75\% and ${\tt PARP}$-${\tt P}$ at 90\% sparsities have lower WERs (higher intelligibility) than its unpruned version. 
For Tacotron2, there is no WER reduction and its WERs remain at $\sim$9\% at up to 40\% sparsity (no change in intelligibility).
Based on (2) and Section~\ref{subsec:exp_naturalness}, we can further conclude that the CNN-based vocoder is highly prunable, with little to no naturalness and intelligibility degradation at up to almost 90\% sparsity.

\vspace{-4mm}
\subsection{Does Sparsity change Prosody?}
\label{subsec:exp_prosody}
\vspace{-6mm}

\begin{figure}[!h]
\centering
\includegraphics[width=0.88\linewidth]{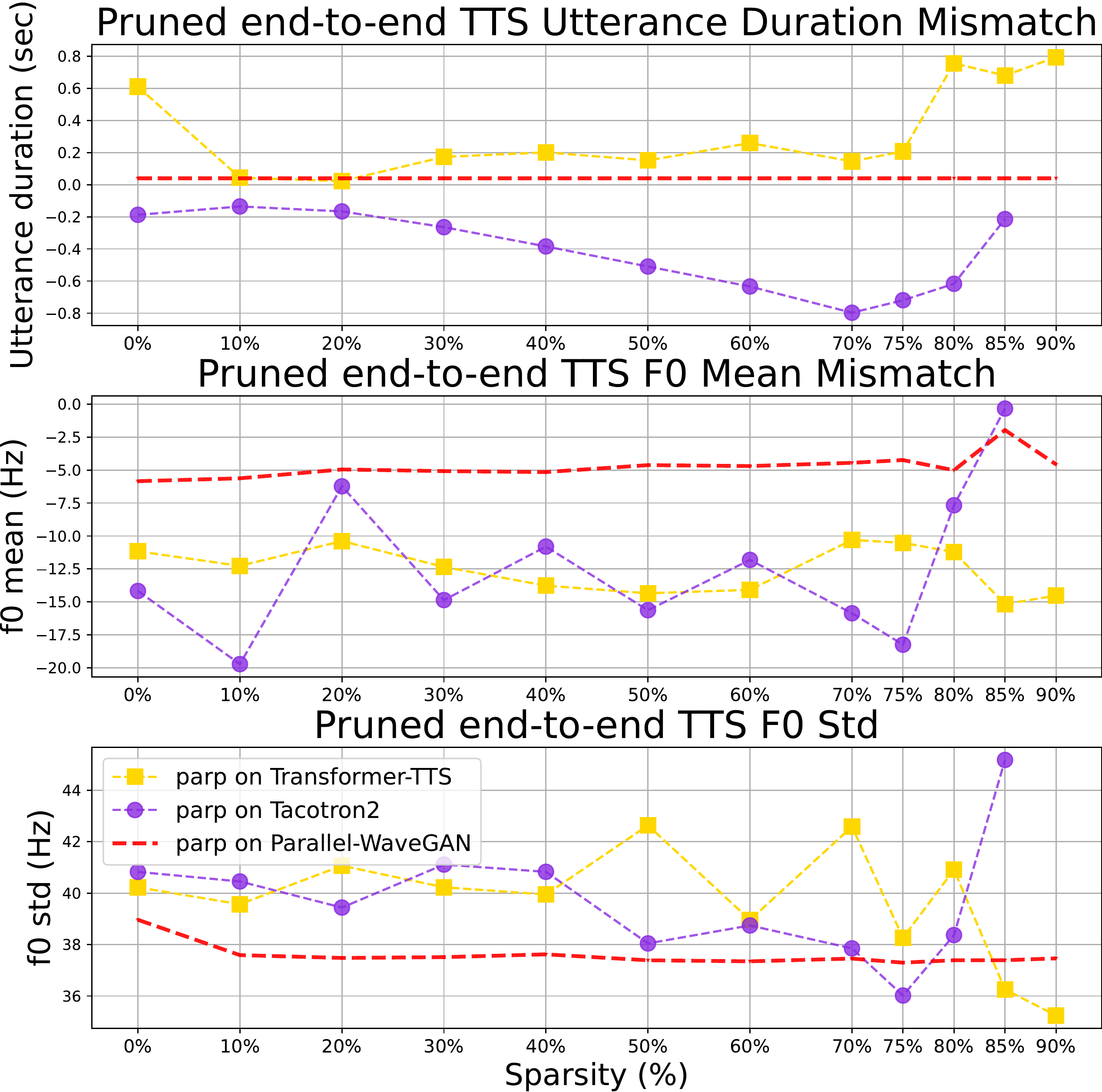}
\vspace{-4mm}
\caption{\textbf{Top} is utterance duration mismatch (in seconds), \textbf{Middle} is $F_{0}$ mean mismatch (in Hz), and \textbf{Bottom} is $F_{0}$ std (in Hz). 
Mismatches are calculated against ground truth recordings. 
Full model (0\%) results are also included. 
}
\label{fig:f0}
\vspace{-3mm}
\end{figure}

We used synthetic speech's utterance duration and mean/std $F_{0}$ across time as three rough proxies for prosody.
Fig~\ref{fig:f0} plots the prosody mismatch between pruned models and ground truth recordings across model combinations. 
Observe ${\tt PARP}$ on Tacotron2 and on Transformer-TTS result in visible differences in prosody changes over sparsities. 
In the top plot, pruned Transformer-TTS (yellow dotted line) have the same utterance duration (+0.2 seconds over ground truth) at 10\%$\sim$75\% sparsities, while in the same region, pruned Tacotron2 (purple dotted line) results in a linear decrease in duration (-0.2$\sim$-0.8 seconds).
Indeed, we confirmed by listening to synthesis samples that pruning Tacotron2 leads to shorter utterance duration as sparsity increases. 

In the middle plot and up to 80\% sparsity, pruned Tacotron2 models have a much large $F_{0}$ mean variation (-20$\sim$-7.5 Hz) compared to that of Transformer-TTS (-10$\sim$-15 Hz). 
We hypothesize that ${\tt PARP}$ on RNN-based models leads to unstable gradients through time during training, while Transformer-based models are easier to prune.
Further, ${\tt PARP}$ on WaveGAN (red dotted line) has a minimal effect on both metrics across sparsities, which leads us to another hypothesis that \textit{vocoder is not responsible for prosody generation}.

In the bottom plot and up to 80\% sparsity, pruned models all have minimal $F_{0}$ std variations ($\le2$ Hz) compared to 53Hz ground truth $F_{0}$ std. 
We infer that at reasonable sparsities, \textit{pruning does not hurt prosodic expressivity}, due to lack of $F_{0}$ oversmoothing~\cite{tan2021survey,zen2009statistical}.

\vspace{-4mm}
\subsection{Does more finetuning data improve sparsity?}
\label{subsec:exp_finetune_data}
\vspace{-2mm}

In ~\cite{lai2021parp}, the authors attain pruned wav2vec 2.0 at much higher sparsity without WER increase given sufficient finetuning data (10h Librispeech split).
Therefore, one question we had was, how much finetuning data is ``good enough" for pruning end-to-end TTS? 
We did two sets of experiments, and for each, we modify the amount of data in ${\tt PARP}$'s Step 2, while keeping $\theta_0$ as is (trained on full LJspeech). 

\begin{figure}[!ht]
\centering
\includegraphics[width=0.875\linewidth]{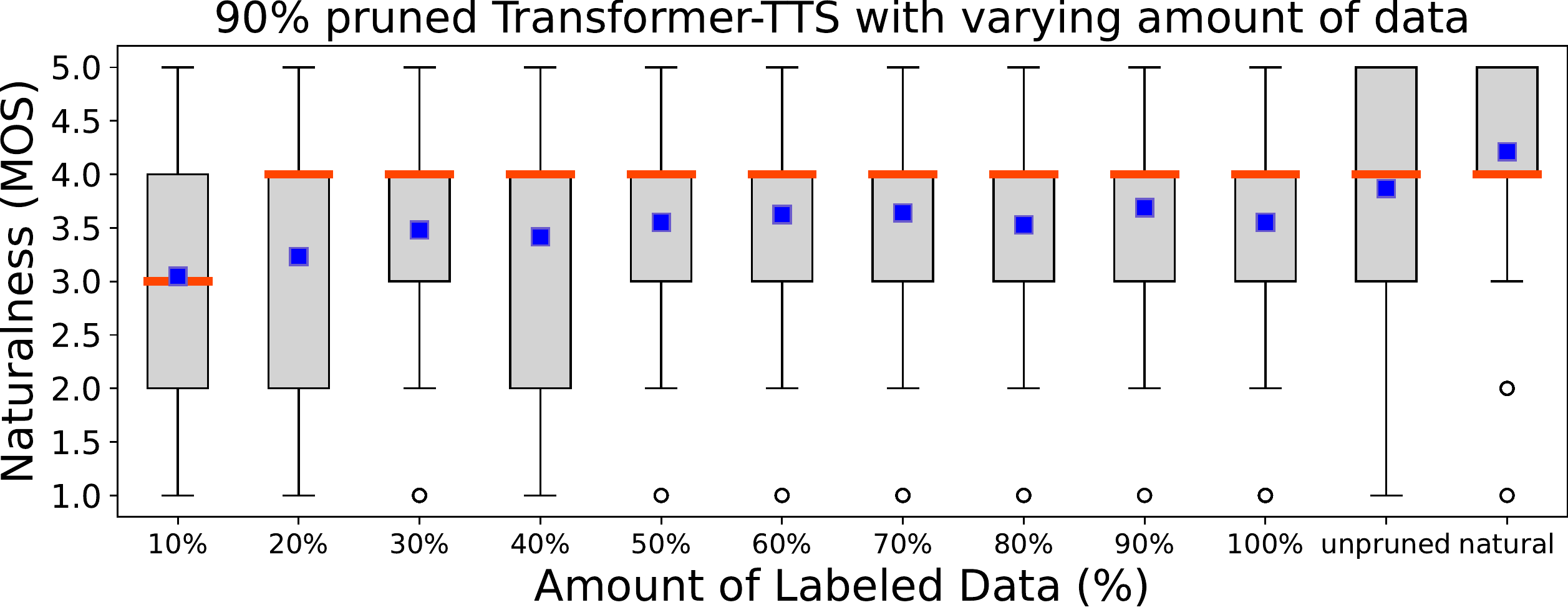}
\vspace{-4mm}
\caption{Effect of amount of finetuning data in ${\tt PARP}$'s Step 2 on MOS score. 
Model is 90\% pruned Transformer-TTS.} 
\label{fig:finetune_data}
\vspace{-4mm}
\end{figure}

The first set of experiments result is Fig~\ref{fig:finetune_data}.
Even at as high as 90\% sparsity, 30\% of finetuning data ($\sim$7.2h) is enough for ${\tt PARP}$ to reach the same level of naturalness as full data\footnote{The effect of using less data to obtain $\theta_0$ remains unclear.}.
The other set of experiment is TTS-Augmentation for utilizing additional unspoken text ($\sim$100h, no domain mismatch) for ${\tt PARP}$'s Step 2. 
In Fig~\ref{fig:wer}'s bottom plot, we see TTS-Augmentations (dark \& light green lines) bear minimal effect on the synthetic speech WERs.
However, Table~\ref{tab:ab_results} indicates that TTS-Augmentation ${\tt PARP}$+${\tt seq}$-${\tt aug}$ does statistically improve ${\tt PARP}$ in naturalness and intelligibility subjective testings. 
\vspace{-3mm}

\vspace{-3.75mm}
\subsection{Ablations}
\label{subsec:exp_ablation}
\vspace{-3mm}
\myparagraph{Knowledge Distillation hurts ${\tt PARP}$}
Surprisingly, we found combining knowledge distillation from teacher model $f(\theta_{D})$ with ${\tt PARP}$ significantly reduces the synthesis quality, see ${\tt PARP}$+${\tt KD}$ v.s. ${\tt PARP}$ in Table~\ref{tab:ab_results}.
Perhaps more careful tuning is required to make ${\tt KD}$ work. 

\vspace{-1mm}
\myparagraph{Importance of $\theta_0$}
Bottom plot of Fig~\ref{fig:wer} (black dotted line) and Table~\ref{tab:ab_results} (${\tt PARP}$ v.s. ${\tt PARP}$-${\tt RI}$) demonstrate the importance of setting the initial model weight $\theta_0$.
In both cases, we set $\theta_0$ to random initialization (${\tt RI}$) instead of $\theta_D$ on LJspeech.

\vspace{-1mm}
\myparagraph{Effectiveness of ${\tt IMP}$}
Table~\ref{tab:ab_results} shows the clear advantage of ${\tt PARP}$-${\tt P}$ over ${\tt IMP}$ at high sparsities, yet ${\tt PARP}$ is not strictly better than ${\tt IMP}$. 

\vspace{-3mm}
\begin{table}[!h]
\vspace{-3mm}
\caption{A/B testing results. 
Each comparison is over 200 HITs. 
\textbf{Bold} numbers are statistical significant under pairwise z test.}
\vspace{1mm}
\label{tab:ab_results}
\centering 
\resizebox{0.8\linewidth}{!}{
\begin{tabular}{llcrr}
\toprule
\multirow{2}{*}{Proposal} & \multirow{2}{*}{Baseline} & Sparsity & \multicolumn{2}{c}{Preference over Baseline} \\
\cmidrule(lr){4-5}
{} & {} & {Level} & {Naturalness} & {Intelligibility}\\
\midrule
\midrule
\multicolumn{5}{l}{\textbf{pruned Transformer-TTS + unpruned Parallel WaveGAN}}\\
${\tt PARP}$-${\tt P}$ & ${\tt PARP}$   & 90\%      & \bf{57\%} & \bf{66\%} \\
                        &               & 95\%      & \bf{63\%} & \bf{64\%}\\
${\tt PARP}$+${\tt KD}$ & ${\tt PARP}$  & 70\%      & \bf{40\%} & \bf{43\%}\\        
                        &               & 90\%      & \bf{36\%} & \bf{27\%} \\
${\tt PARP}$-${\tt P}$  & ${\tt IMP}$   & 90\%      & 53\%      & 51\% \\
                        &               & 95\%      & \bf{64\%} &  \bf{61\%} \\
${\tt PARP}$  & ${\tt IMP}$ & 30\%    & 54\%         & \bf{58\%}  \\
              &             & 50\%    & 46\%         &  54\% \\
              &             & 90\%    & \bf{42\%}    &  \bf{37\%} \\
${\tt PARP}$  & ${\tt PARP}$-${\tt RI}$ & 10\%    & 55\%       &  \bf{57\%}  \\              
              &                 & 30\%    & 55\%       &  53\% \\      
              &                 & 50\%    & \bf{56\%}  &  \bf{67\%}  \\      
              &                 & 70\%    & 53\%       &  53\% \\      
              &                 & 90\%    & \bf{60\%}  &  \bf{56\%} \\      
${\tt PARP}$+${\tt seq}$-${\tt aug}$  & ${\tt PARP}$ & 10\%    & \bf{58\%}       &  \bf{58\%} \\    
          &                 & 30\%    & 52\%       &  \bf{57\%} \\      
          &                 & 50\%    & \bf{44\%}       &  \bf{41\%} \\      
          &                 & 70\%    & \bf{57\%}       &  54\% \\      
          &                 & 90\%    & 51\%       &  \bf{56\%} \\      
\bottomrule
\end{tabular}}
\vspace{-5mm}
\end{table}

\vspace{-2mm}
\section{Remarks}
\label{sec:conclusions}
\vspace{-3.5mm}
\myparagraph{Significance}
This work is scientific in nature, as most of our results arose from analyzing pruned models via large-scale experimentation and testing.  
In fact, we are less interested in answering questions like ``how much can Tacotron2 be reduced to?", and more curious about inquiries along the line of, ``what are patterns/properties unique to end-to-end TTS?" 
Continuing in this direction should allow us to understand more in depth how end-to-end TTS models are trained.

    
    
    
    
\vspace{-1mm}
\myparagraph{Future Work} Possible extensions upon this work are 1) inclusion of an ASR loss term in PARP's step 2 for melspec pruning, 2) multi-lingual/multi-speaker TTS pruning, 3) further study on \textit{why} over-parameterization benefit end-to-end TTS, 4) showing the Lottery Ticket Hypothesis~\cite{frankle2018lottery} exists in end-to-end TTS, and 5) determining if insights from pruning help design better TTS model architectures.





\vfill\pagebreak

{\footnotesize
\bibliographystyle{IEEEbib}
\bibliography{main}}

\newpage
\onecolumn
\section{Supplementary Materials}
\label{sec:appendix}

\subsection{Philosophy}
We clarify here again the philosophy of this piece of work -- we are interested in quantifying the effects of sparsity on speech perceptions (measured via naturalness, intelligibility, and prosody), as there has not been any systematic study in this direction. 
The pruning methods, either ${\tt PARP}$ or ${\tt IMP}$, are merely engineering ``tools" to probe the underlying phenomenon. 
It is not too difficult to get higher sparsities or introduce fancier pruning techniques as noted in this study, yet they are less scientifically interesting at this moment.

\subsection{${\tt IMP}$ Implementation}
Normally, ${\tt IMP}$s are iterated many times depending on the training costs. 
In iteration $i$, the ${\tt IMP}$ mask from the previous iteration $m_{i-1}$ is fixed, and only the remaining weights are considered in ${\tt UMP}$.
In other words, each iteration's sparsity pattern is built on top of the that from the previous iteration.
For simplicity, our ${\tt IMP}$ implementation only has 1 iteration. 
In some literature, this is termed One-Shot Magnitude Pruning ${\tt OMP}$, though it is also not exactly our case due to our starting point is a fully trained model $\theta_D$ instead of a randomly initialized model. 

\subsection{TTS-Augmentation Implementations}
We experimented with two TTS-Augmentation policies for ${\tt PARP}$ with initial model $f(\theta_D)$ trained on LJspeech $\mathcal{D}$ and additional unspoken text from Librispeech $\bm{X}_u$: 
\begin{itemize}
    \item ${\tt PARP}$+${\tt seq}$-${\tt aug}$: synthesize $\bm{X}_u$ with $f(\theta_D)$ to obtain $\mathcal{D}_u = (\bm{X}_u, f(\bm{X}_u;\theta_D))$. 
    First, perform ${\tt PARP}$'s Step 2 at target sparsity on $\mathcal{D}_u$ with initial model weight $\theta_D$.
    Obtain a pruned model with model weight $m_{U} \odot \theta_U^*$. 
    Next, perform ${\tt PARP}$'s Step 2 at target sparsity on $\mathcal{D}$ with initial model weight $m_{U} \odot \theta_U^*$.
    Return $m_{D} \odot \theta_D^*$.

    \item ${\tt PARP}$+${\tt mix}$-${\tt aug}$: synthesize $\bm{X}_u$ with $f(\theta_D)$ to obtain $\mathcal{D}_u = (\bm{X}_u, f(\bm{X}_u;\theta_D))$. 
    Directly perform ${\tt PARP}$'s Step 2 on $(\mathcal{D}, \mathcal{D}_u)$ with initial model weight $\theta_D$.
    Return $m_{D} \odot \theta_D^*$.
\end{itemize}
Table~\ref{tab:tts_aug_ab_results} is the subjective A/B testing results comparing the two augmentation policies at different sparsities. 
Although there is no statistical preference between the two, we note that pruned models with different augmentation policies do subjectively sound different from one another. 

\vspace{-2mm}
\begin{table}[!h]
\caption{TTS-Augmentation A/B testing results. 
Each comparison is over 200 HITs.}
\vspace{1mm}
\label{tab:tts_aug_ab_results}
\centering 
\resizebox{0.6\linewidth}{!}{
\begin{tabular}{llcrr}
\toprule
\multirow{2}{*}{Proposal} & \multirow{2}{*}{Baseline} & Sparsity & \multicolumn{2}{c}{Preference over Baseline} \\
\cmidrule(lr){4-5}
{} & {} & {Level} & {Naturalness} & {Intelligibility}\\
\midrule
\midrule
\multicolumn{5}{l}{\textbf{pruned Transformer-TTS + unpruned Parallel WaveGAN}}\\
${\tt PARP}$+${\tt seq}$-${\tt aug}$  & ${\tt PARP}$+${\tt mix}$-${\tt aug}$ & 10\%    & 53\%       &  - \\ 
          &                 & 30\%    & 55\%       &  - \\      
          &                 & 50\%    & 46\%       &  - \\      
          &                 & 70\%    & 47\%       &  - \\      
          &                 & 90\%    & 53\%       &  - \\      
\bottomrule
\end{tabular}}
\vspace{-4mm}
\end{table}

\newpage
\subsection{Statistical Testing Result}
Below are the results of running Mann-Whitney U test on our Naturalness MOS tests at a significance level $p\le 0.05$, where $\bullet$ indicates pair-wise statistically significant, and $\square$ indicates the opposite.
The order in $x$ and $y$ axes are shuffled due to our MOS test implementation.
Cross reference with Figure~\ref{fig:mos}, we have the following conclusions: 
\begin{itemize}
    \item ${\tt PARP}$ on Transformer-TTS + unpruned Parallel WaveGAN (Table~\ref{tab:mann_whitney_test_1}): no pruned model is statistically better than the unpruned model, and pruned model is only statistically worse starting at 90\% sparsity.
    
    \item ${\tt PARP}$ on Tacotron2 + unpruned Parallel WaveGAN (Table~\ref{tab:mann_whitney_test_2}): 30\% pruned model is statistically better than the unpruned model, and pruned model is only statistically worse starting at 90\% sparsity.
    
    \item Unpruned Transformer-TTS + ${\tt PARP}$ on Parallel WaveGAN (Table~\ref{tab:mann_whitney_test_3}): no pruned model is statistically better than the unpruned model, and no pruned model model is statistically worse either. 
    
    \item Unpruned Tacotron2 + ${\tt PARP}$ on Parallel WaveGAN (Table~\ref{tab:mann_whitney_test_4}): no pruned model is statistically better than the unpruned model, and pruned model is only statistically worse starting at 85\% sparsity.
\end{itemize}
Based on the above statistical testings, we are \textit{not} claiming that pruning definitely improves TTS training. 
Instead, at the right conditions, pruned models \textit{could} perform better than the unpruned ones.  

\begin{table}[!h]
\vspace{-3mm}
\caption{${\tt PARP}$ on Transformer-TTS + unpruned Parallel WaveGAN}
\vspace{2mm}
\label{tab:mann_whitney_test_1}
\centering 
\resizebox{0.9\linewidth}{!}{
\begin{tabular}{c|c|c|c|c|c|c|c|c|c|c|c|c|c|c|c|c}
 & natural  & 70\%  & 30\%  & natural  & 99\%  & 90\%  & 50\%  & 95\%  & full  & 80\%  & 75\%  & 40\%  & 85\%  & 10\%  & 20\%  & 60\% \\ \hline
natural & - &  &  &  &  &  &  &  &  &  &  &  &  &  &  & \\ \hline
70\% & $\bullet$  & - &  &  &  &  &  &  &  &  &  &  &  &  &  & \\ \hline
30\% & $\bullet$  & $\bullet$  & - &  &  &  &  &  &  &  &  &  &  &  &  & \\ \hline
natural & $\square$  & $\bullet$  & $\square$  & - &  &  &  &  &  &  &  &  &  &  &  & \\ \hline
99\% & $\bullet$  & $\bullet$  & $\bullet$  & $\bullet$  & - &  &  &  &  &  &  &  &  &  &  & \\ \hline
90\% & $\bullet$  & $\square$  & $\bullet$  & $\bullet$  & $\bullet$  & - &  &  &  &  &  &  &  &  &  & \\ \hline
50\% & $\bullet$  & $\bullet$  & $\square$  & $\square$  & $\bullet$  & $\bullet$  & - &  &  &  &  &  &  &  &  & \\ \hline
95\% & $\bullet$  & $\bullet$  & $\bullet$  & $\bullet$  & $\square$  & $\bullet$  & $\bullet$  & - &  &  &  &  &  &  &  & \\ \hline
full & $\bullet$  & $\square$  & $\square$  & $\bullet$  & $\bullet$  & $\bullet$  & $\square$  & $\bullet$  & - &  &  &  &  &  &  & \\ \hline
80\% & $\bullet$  & $\square$  & $\bullet$  & $\bullet$  & $\bullet$  & $\square$  & $\bullet$  & $\bullet$  & $\bullet$  & - &  &  &  &  &  & \\ \hline
75\% & $\bullet$  & $\square$  & $\square$  & $\bullet$  & $\bullet$  & $\bullet$  & $\square$  & $\bullet$  & $\square$  & $\bullet$  & - &  &  &  &  & \\ \hline
40\% & $\bullet$  & $\bullet$  & $\square$  & $\square$  & $\bullet$  & $\bullet$  & $\square$  & $\bullet$  & $\square$  & $\bullet$  & $\square$  & - &  &  &  & \\ \hline
85\% & $\bullet$  & $\square$  & $\bullet$  & $\bullet$  & $\bullet$  & $\square$  & $\bullet$  & $\bullet$  & $\square$  & $\square$  & $\bullet$  & $\bullet$  & - &  &  & \\ \hline
10\% & $\bullet$  & $\bullet$  & $\square$  & $\square$  & $\bullet$  & $\bullet$  & $\square$  & $\bullet$  & $\square$  & $\bullet$  & $\square$  & $\square$  & $\bullet$  & - &  & \\ \hline
20\% & $\bullet$  & $\square$  & $\square$  & $\bullet$  & $\bullet$  & $\bullet$  & $\square$  & $\bullet$  & $\square$  & $\bullet$  & $\square$  & $\square$  & $\square$  & $\square$  & - & \\ \hline
60\% & $\bullet$  & $\square$  & $\square$  & $\bullet$  & $\bullet$  & $\bullet$  & $\square$  & $\bullet$  & $\square$  & $\bullet$  & $\square$  & $\square$  & $\bullet$  & $\square$  & $\square$  & -\\ \hline
\end{tabular}}
\end{table}

\begin{table*}[!h]
\vspace{-3mm}
\caption{${\tt PARP}$ on Tacotron2 + unpruned Parallel WaveGAN}
\vspace{2mm}
\label{tab:mann_whitney_test_2}
\centering 
\resizebox{0.9\linewidth}{!}{
\begin{tabular}{c|c|c|c|c|c|c|c|c|c|c|c|c|c|c|c|c}
 & natural  & 70\%  & 30\%  & natural  & 99\%  & 90\%  & 50\%  & 95\%  & full  & 80\%  & 75\%  & 40\%  & 85\%  & 10\%  & 20\%  & 60\% \\ \hline
natural & - &  &  &  &  &  &  &  &  &  &  &  &  &  &  & \\ \hline
70\% & $\bullet$  & - &  &  &  &  &  &  &  &  &  &  &  &  &  & \\ \hline
30\% & $\square$  & $\square$  & - &  &  &  &  &  &  &  &  &  &  &  &  & \\ \hline
natural & $\square$  & $\bullet$  & $\square$  & - &  &  &  &  &  &  &  &  &  &  &  & \\ \hline
99\% & $\bullet$  & $\bullet$  & $\bullet$  & $\bullet$  & - &  &  &  &  &  &  &  &  &  &  & \\ \hline
90\% & $\bullet$  & $\bullet$  & $\bullet$  & $\bullet$  & $\square$  & - &  &  &  &  &  &  &  &  &  & \\ \hline
50\% & $\bullet$  & $\square$  & $\square$  & $\bullet$  & $\bullet$  & $\bullet$  & - &  &  &  &  &  &  &  &  & \\ \hline
95\% & $\bullet$  & $\bullet$  & $\bullet$  & $\bullet$  & $\square$  & $\square$  & $\bullet$  & - &  &  &  &  &  &  &  & \\ \hline
full & $\bullet$  & $\square$  & $\bullet$  & $\bullet$  & $\bullet$  & $\bullet$  & $\square$  & $\bullet$  & - &  &  &  &  &  &  & \\ \hline
80\% & $\bullet$  & $\square$  & $\bullet$  & $\bullet$  & $\bullet$  & $\bullet$  & $\square$  & $\bullet$  & $\square$  & - &  &  &  &  &  & \\ \hline
75\% & $\bullet$  & $\square$  & $\square$  & $\bullet$  & $\bullet$  & $\bullet$  & $\square$  & $\bullet$  & $\square$  & $\bullet$  & - &  &  &  &  & \\ \hline
40\% & $\square$  & $\square$  & $\square$  & $\square$  & $\bullet$  & $\bullet$  & $\square$  & $\bullet$  & $\square$  & $\square$  & $\square$  & - &  &  &  & \\ \hline
85\% & $\bullet$  & $\square$  & $\bullet$  & $\bullet$  & $\bullet$  & $\bullet$  & $\square$  & $\bullet$  & $\square$  & $\square$  & $\square$  & $\square$  & - &  &  & \\ \hline
10\% & $\bullet$  & $\square$  & $\square$  & $\bullet$  & $\bullet$  & $\bullet$  & $\square$  & $\bullet$  & $\square$  & $\square$  & $\square$  & $\square$  & $\square$  & - &  & \\ \hline
20\% & $\bullet$  & $\square$  & $\square$  & $\bullet$  & $\bullet$  & $\bullet$  & $\square$  & $\bullet$  & $\square$  & $\square$  & $\square$  & $\square$  & $\square$  & $\square$  & - & \\ \hline
60\% & $\bullet$  & $\square$  & $\square$  & $\bullet$  & $\bullet$  & $\bullet$  & $\square$  & $\bullet$  & $\square$  & $\square$  & $\square$  & $\square$  & $\square$  & $\square$  & $\square$  & -\\ \hline
\end{tabular}}
\end{table*}

\begin{table*}[!h]
\vspace{-3mm}
\caption{Unpruned Transformer-TTS + ${\tt PARP}$ on Parallel WaveGAN}
\vspace{2mm}
\label{tab:mann_whitney_test_3}
\centering 
\resizebox{0.9\linewidth}{!}{
\begin{tabular}{c|c|c|c|c|c|c|c|c|c|c|c|c|c|c|c}
 & natural  & 60\%  & 20\%  & natural  & 85\%  & 40\%  & 88\%  & full  & 75\%  & 70\%  & 30\%  & 80\%  & full  & 10\%  & 50\% \\ \hline
natural & - &  &  &  &  &  &  &  &  &  &  &  &  &  & \\ \hline
60\% & $\bullet$  & - &  &  &  &  &  &  &  &  &  &  &  &  & \\ \hline
20\% & $\bullet$  & $\square$  & - &  &  &  &  &  &  &  &  &  &  &  & \\ \hline
natural & $\square$  & $\bullet$  & $\bullet$  & - &  &  &  &  &  &  &  &  &  &  & \\ \hline
85\% & $\bullet$  & $\square$  & $\square$  & $\bullet$  & - &  &  &  &  &  &  &  &  &  & \\ \hline
40\% & $\bullet$  & $\square$  & $\square$  & $\bullet$  & $\square$  & - &  &  &  &  &  &  &  &  & \\ \hline
88\% & $\bullet$  & $\square$  & $\square$  & $\bullet$  & $\square$  & $\square$  & - &  &  &  &  &  &  &  & \\ \hline
full & $\bullet$  & $\square$  & $\square$  & $\bullet$  & $\square$  & $\square$  & $\square$  & - &  &  &  &  &  &  & \\ \hline
75\% & $\bullet$  & $\square$  & $\bullet$  & $\bullet$  & $\square$  & $\square$  & $\square$  & $\square$  & - &  &  &  &  &  & \\ \hline
70\% & $\bullet$  & $\square$  & $\bullet$  & $\bullet$  & $\square$  & $\square$  & $\square$  & $\square$  & $\square$  & - &  &  &  &  & \\ \hline
30\% & $\bullet$  & $\square$  & $\square$  & $\bullet$  & $\square$  & $\square$  & $\square$  & $\square$  & $\square$  & $\square$  & - &  &  &  & \\ \hline
80\% & $\bullet$  & $\square$  & $\square$  & $\bullet$  & $\square$  & $\square$  & $\square$  & $\square$  & $\square$  & $\square$  & $\square$  & - &  &  & \\ \hline
full & $\bullet$  & $\square$  & $\square$  & $\bullet$  & $\square$  & $\square$  & $\square$  & $\square$  & $\square$  & $\square$  & $\square$  & $\square$  & - &  & \\ \hline
10\% & $\bullet$  & $\square$  & $\square$  & $\bullet$  & $\square$  & $\square$  & $\square$  & $\square$  & $\square$  & $\square$  & $\square$  & $\square$  & $\square$  & - & \\ \hline
50\% & $\bullet$  & $\square$  & $\square$  & $\bullet$  & $\square$  & $\square$  & $\square$  & $\square$  & $\square$  & $\square$  & $\square$  & $\square$  & $\square$  & $\square$  & -\\ \hline
\end{tabular}}
\end{table*}

\begin{table*}[!ht]
\vspace{-3mm}
\caption{Unpruned Tacotron2 + ${\tt PARP}$ on Parallel WaveGAN}
\vspace{2mm}
\label{tab:mann_whitney_test_4}
\centering 
\resizebox{0.9\linewidth}{!}{
\begin{tabular}{c|c|c|c|c|c|c|c|c|c|c|c|c|c|c|c}
 & natural  & 60\%  & 20\%  & natural  & 85\%  & 40\%  & 88\%  & full  & 75\%  & 70\%  & 30\%  & 80\%  & full  & 10\%  & 50\% \\ \hline
natural & - &  &  &  &  &  &  &  &  &  &  &  &  &  & \\ \hline
60\% & $\bullet$  & - &  &  &  &  &  &  &  &  &  &  &  &  & \\ \hline
20\% & $\bullet$  & $\square$  & - &  &  &  &  &  &  &  &  &  &  &  & \\ \hline
natural & $\square$  & $\bullet$  & $\bullet$  & - &  &  &  &  &  &  &  &  &  &  & \\ \hline
85\% & $\bullet$  & $\square$  & $\square$  & $\bullet$  & - &  &  &  &  &  &  &  &  &  & \\ \hline
40\% & $\bullet$  & $\square$  & $\square$  & $\bullet$  & $\square$  & - &  &  &  &  &  &  &  &  & \\ \hline
88\% & $\bullet$  & $\bullet$  & $\bullet$  & $\bullet$  & $\square$  & $\square$  & - &  &  &  &  &  &  &  & \\ \hline
full & $\bullet$  & $\square$  & $\square$  & $\bullet$  & $\bullet$  & $\bullet$  & $\bullet$  & - &  &  &  &  &  &  & \\ \hline
75\% & $\bullet$  & $\square$  & $\square$  & $\bullet$  & $\bullet$  & $\square$  & $\bullet$  & $\square$  & - &  &  &  &  &  & \\ \hline
70\% & $\bullet$  & $\square$  & $\square$  & $\bullet$  & $\square$  & $\square$  & $\bullet$  & $\square$  & $\square$  & - &  &  &  &  & \\ \hline
30\% & $\bullet$  & $\square$  & $\square$  & $\bullet$  & $\square$  & $\square$  & $\bullet$  & $\square$  & $\square$  & $\square$  & - &  &  &  & \\ \hline
80\% & $\bullet$  & $\square$  & $\square$  & $\bullet$  & $\square$  & $\square$  & $\bullet$  & $\square$  & $\square$  & $\square$  & $\square$  & - &  &  & \\ \hline
full & $\bullet$  & $\square$  & $\square$  & $\bullet$  & $\bullet$  & $\bullet$  & $\bullet$  & $\square$  & $\square$  & $\square$  & $\square$  & $\square$  & - &  & \\ \hline
10\% & $\bullet$  & $\square$  & $\square$  & $\bullet$  & $\square$  & $\square$  & $\bullet$  & $\square$  & $\square$  & $\square$  & $\square$  & $\square$  & $\square$  & - & \\ \hline
50\% & $\bullet$  & $\square$  & $\square$  & $\bullet$  & $\bullet$  & $\square$  & $\bullet$  & $\square$  & $\square$  & $\square$  & $\square$  & $\square$  & $\square$  & $\square$  & -\\ \hline
\end{tabular}}
\end{table*}

\newpage~\newpage
\subsection{AMT Testing User Interface}
We append our sample AMT user interfaces used in our study for naturalness MOS testing (Figure~\ref{fig:mos} and Figure~\ref{fig:finetune_data}) and naturalness \& intelligibility A/B testing (Table~\ref{tab:ab_results} and Table~\ref{tab:tts_aug_ab_results}). 
For MOS tests, reference natural recordings are included in the instruction. 
About \$2800 are spent on AMT testing. 

\begin{figure*} [!h]
\includegraphics[width=\linewidth]{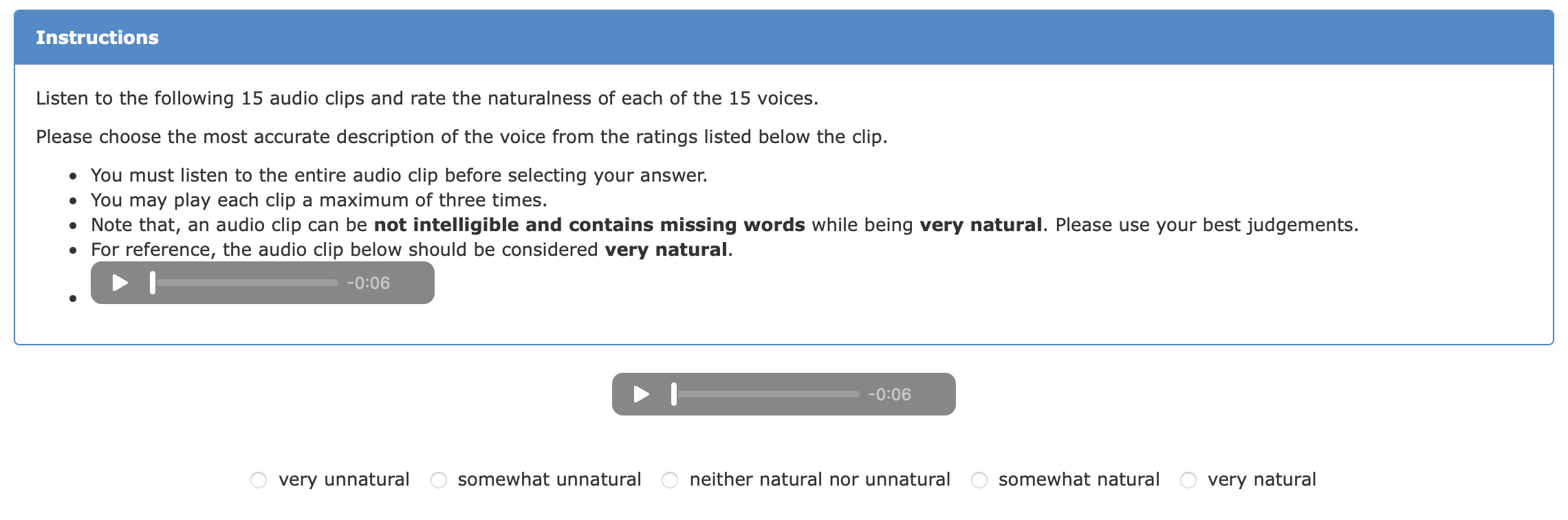}
\centering
\caption{AMT user interface for the 5-point (1-point increment) scale MOS naturalness testing.}
\label{fig:amt_natrual_mos}
\end{figure*}

\begin{figure*} [!h]
\includegraphics[width=\linewidth]{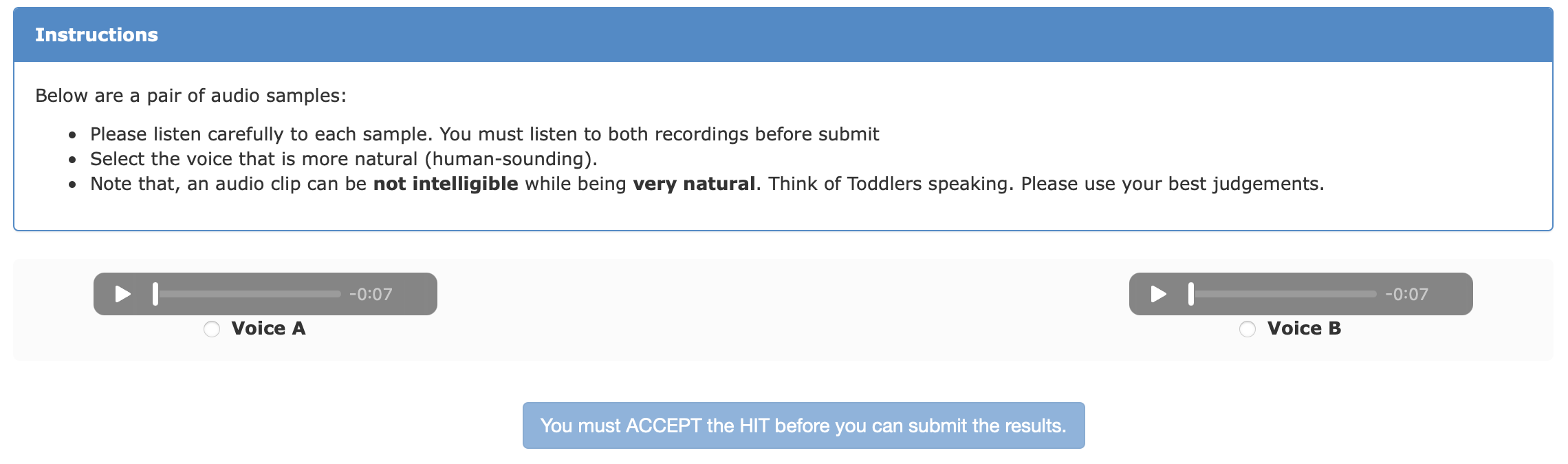}
\centering
\caption{AMT user interface for A/B testing for naturalness.}
\label{fig:amt_natrual_ab}
\end{figure*}

\begin{figure*} [!h]
\includegraphics[width=\linewidth]{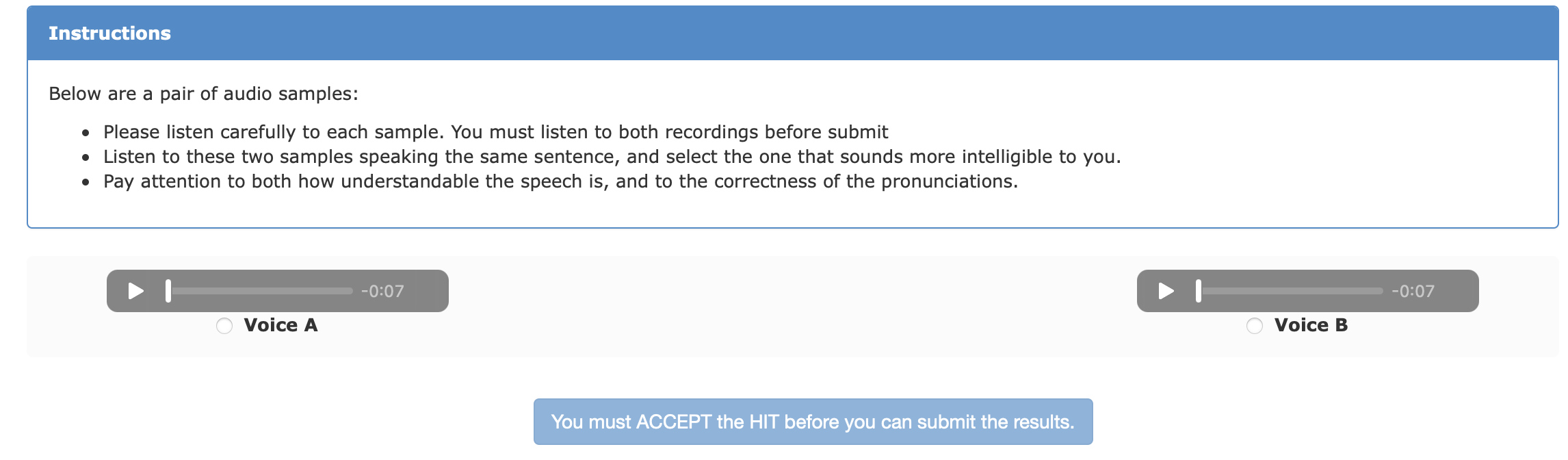}
\centering
\caption{AMT user interface for A/B testing for intelligibility.}
\label{fig:amt_intell_ab}
\end{figure*}

\end{document}